\documentclass[12pt]{article}
\makeatletter
{\endgroup\end{definitionth}\vspace{0.3cm}}
\newtheorem{definitionth}{Definition}
{\endgroup\end{remarkth}}
\newtheorem{remarkth}{Remark}
{\endgroup\end{remarksth}}
\newtheorem{remarksth}[remarkth]{Remarks}
\newtheorem{theorem}{Theorem}
\newtheorem{lemma}{Lemma}
\let\@xpar=\par
\newenvironment{proof}{{\vspace{1ex}\@xpar\noindent}%
{\bf Proof.}\hspace{3mm}\begin{em} }{\penalty10000\end{em}\penalty10000%
\hfill$\Box$\vspace{.5ex}\@xpar\penalty-8000\noindent}

\makeatother
  \usepackage{multirow}
  \usepackage{amsmath}
  \usepackage{amssymb}
  \usepackage{latexsym}
  \usepackage{booktabs}
  \usepackage{amssymb}
  \usepackage{graphicx}
  \usepackage{footnote}
  \usepackage[usenames,dvipsnames]{color}
  \usepackage{url}
  \usepackage{array}
  \usepackage{xcolor}
  \usepackage{amsmath}
  \usepackage{threeparttable, tablefootnote}
  \usepackage{appendix}
  \usepackage{enumitem}
  \usepackage{float}
\usepackage{algorithm}
\usepackage[noend]{algpseudocode}

\def\row#1#2{{#1}_1,{#1}_2,\ldots ,{#1}_{#2}}

\makeatletter
\def\BState{\State\hskip-\ALG@thistlm}
\makeatother

  \usepackage[usenames,dvipsnames]{color}

  \newcolumntype{L}[1]{>{\raggedright\let\newline\\\arraybackslash\hspace{0pt}}m{#1}}
  \newcolumntype{C}[1]{>{\centering\let\newline\\\arraybackslash\hspace{0pt}}m{#1}}
  \newcolumntype{R}[1]{>{\raggedleft\let\newline\\\arraybackslash\hspace{0pt}}m{#1}}

  \newcommand{\except}[1]%
  {{\displaystyle \mathop{{\displaystyle}{\mathop{\ldots}^{\vee}}}^{#1}}}

\begin{document}
  \title{Realistic versus Rational Secret Sharing\thanks{This is a preliminary
    version of a paper accepted for GameSec 2019.}}
  
\author{Yvo Desmedt\thanks{Yvo Desmedt thanks the Jonsson
    Endowment. He is also an Honorary Professor at University College London.}\\
Department of Computer Science,\\
University of Texas at Dallas\\
USA\\
  \and Arkadii Slinko\\
  Department of Mathematics\\
  University of Auckland\\
  New Zealand}


\maketitle

\begin{abstract}
The study of Rational Secret Sharing initiated by Halpern and Teague
  regards the reconstruction of the secret
in secret sharing as a game. It was shown that participants (parties) may refuse to reveal their shares and so the reconstruction may fail. Moreover, a refusal to reveal the share may be a dominant strategy of a party.

In this paper we consider secret sharing as a sub-action or subgame of a larger action/game where the secret opens a possibility of consumption of a certain common good. We claim that utilities of participants will be dependent on the nature of this common good. In particular, Halpern and Teague scenario corresponds to a rivalrous and excludable common good.  We consider the case when this common good is non-rivalrous and non-excludable and find many natural Nash equilibria. We list several applications of secret sharing to demonstrate
our claim and give corresponding scenarios. In such circumstances the secret sharing scheme facilitates a power sharing agreement in the society.  We also state that non-reconstruction may be beneficial for this society and give several examples.

\end{abstract}

\section{Introduction}\label{sec-intro}
The classical solution of backing up data, called the {\it secret}, is to make
a copy. This has the disadvantage that when the trust in the storage facility
is compromised the owner loses the privacy. The mechanical approach of using
several locks to a mechanical safe was known for a long time to researchers in
combinatorics (see e.g., Liu~\cite{Liu68}). The combinatorial solution to the
problem, being inefficient was the inspiration 40 years ago to Shamir~\cite{Shamir79}
to use interpolation of polynomials to obtain a purely
algebraic  solution, which, as was pointed out later~\cite{McElieceSarwate81}, was equivalent to the use of the Reed-Solomon
error-correction code as an erasure code. The owner of the data, called the
{\it dealer\/}, will give shares of the secret to partially trusted parties. In contrast to
Blakley~\cite{Blakley79}, Shamir's approach was {\it perfect}, i.e.,
unauthorised coalitions of users would have no information, better than guessing, about the
original secret. In the last 40 years we have seen an explosion
of applications of secret sharing, which are briefly surveyed in
Section~\ref{sec:apps}.

Halpern and Teague~\cite{Halpern-Teague:2004} considered the reconstruction of the secret from
its shares as a {\it game} endowing the users with greedy utilities. If some participants reveal their shares to others and others do not, the latter may be in a position to recover the secret without those who revealed. They show that the only Nash Equilibrium in such a game is that 
participants do  {\it not\/}  reveal their shares. 
The scenario they considered is, figuratively speaking, the one
in which the secret corresponds to a treasure map to which several pirates have partial information about.  And those who learn the secret will share the treasure between themselves. 



In this paper we will introduce new aspects (and scenarios) related to the reconstruction/use of the secret. We in
particular argue that:
\begin{itemize}
\item There is a variety of economic situations where secret sharing is embedded as a subgame;
\item The secret usually constitutes a common good (or opens the way of producing or consuming of a common good);
\item The utilities of participants and, hence, the characteristics of the reconstruction game will depend 
on the properties of the common good;
\item In particular, in some scenarios the reconstruction game will have a set of natural Nash equilibria with a number of parties revealing their shares leading to the secret recovery.
\end{itemize}

We will now survey some of the modern applications of secret sharing.  We will
then use these to illustrate that in realistic situations the reconstruction of the secret 
corresponds to a subgame or sub-action of a bigger game or bigger action. We will refer to these cases as {\it Realistic 
Secret Sharing}. In particular, we show that secret sharing can serve as a power-sharing agreement in the society. We will revisit the utility
aspects from the above perspectives. 

In this paper we do not consider secret sharing as a repeated game. In such a case cooperation in the reconstruction stage can be reinforced by the introduction of a reputation value \cite{socio-rational}. 
However, as we discuss further, not participation in the reconstruction of the
secret may be caused by reasons other than being selfish and in sucg a case it may be beneficial for the society. 

\section{Modern applications of secret sharing}\label{sec:apps}
Cryptographers have developed a wide range of applications of secret sharing,
which we will now briefly survey. 
\paragraph{\bf Secure Multiparty Computation.}
Secure Multiparty Computation (MPC)\footnote{Sometimes the acronym SMC is
  used.} introduced by Yao~\cite{Yao86} (see also~\cite{Yao82a}),
is one of the first cryptographic uses of secret sharing.
Secure Multiparty Computation allows $n$ parties to compute a function in $n$
inputs (one per participant) without leaking any extra information about the 
inputs. With extra we mean something that does not logically follow from the
output. Yao's result was restricted to $n=2$ and made use of 
an unproven cryptographic assumption. In full, the problem
was solved using secret sharing in~\cite{BenOrGoldwasserWigderson88}.
Several follow-up papers have also used secret sharing in this context,
e.g.,~\cite{GennaroRabinRabin98,Beerliova-TrubiniovaHirth08}.

\paragraph{\bf Threshold cryptography.}
In Threshold Cryptography the secret key of a cryptosystem is
shared between a group of participants~\cite{Boyd86,CroftHarris86,Desmedt87}.  The topic of threshold
decryption (see e.g.,~\cite{DesmedtFrankel89Crypto}) and threshold signatures
have received the most attention.

In threshold decryption, the secret key used to decrypt ciphertexts
(encrypted messages),
is shared. In some scenarios one does not trust the receiver's decryption device.
Using multiple decryption units and threshold decryption solves this trust issue.

Another application of threshold decryption occurs when the message
is intended for a group of people~\cite{Desmedt87} and not for
a single individual. An example of such a group might be a parliament,
or a board of directors. It is assumed the organisation as a whole
has its own public key.

Threshold signatures prevent fraud by disabling a single party, or in general
an unauthorised subset of insiders, from signing in the name of their organisation.
The organisation could be any legal body that wants to act as
a single entity having its own public key.

\paragraph{\bf Perfectly Secure Message Transmission.}
The problem of using an unreliable network for reliable communication is quite
old and might have been the inspiration to study connectivity of networks.

The first contribution by cryptographers in this area goes back to
1993~\cite{DolevDworkWaartsYung93}. Dolev et al.\ considered that the sender and
the receiver do not trust unproven cryptographic assumptions (so no public key
cryptography)
and do not share a one-time pad. Despite these limitations they still want to
communicate 
reliably and privately. Dolev et al.\ considered that $t$ {\it nodes\/} in a
network are untrustworthy and wondered how to communicate in such circumstances. They considered the following two
scenarios: the network has:
\begin{enumerate}
\item\label{enum:one-way} only one-way communication, or
\item\label{enum:two-way} only two-way communications.
\end{enumerate}  
Dolev et al.\ gave an efficient solution for the first case
by using $(t+1)$-out-of-$(3t+1)$
secret sharing and assuming a $(3t+1)$-connected network. Each of the $3t+1$
disjoint paths between sender and receiver is used to send a share. One had to wait 25
years before an efficient solution was found for the 2nd
setting~\cite{KurosawaSuzuki09}. Their solution also uses secret sharing.

For surveys (partially outdated) consult~\cite{Desmedt06BT,Desmedt05ITW}.

\paragraph{\bf Multi Cloud.}
In 2010 IBM considered the issue that cloud servers might not be trusted, from
both a reliability and a privacy perspective~\cite{mishra2013cloud}. Instead of having a
single cloud server, they use multiple cloud servers each receiving shares of
the secret.

\section{Economics underpinnings: common good and individual utility}

There are different models of how agents may extract some utility knowing the secret. As is usual in economics, the utility comes from some sort of consumption of a certain good. Since the access to the consumption comes from collaborating with others (an authorised coalition must be formed) the good to be consumed falls under the category of common good.

The concept of the common good can be traced back to ancient Greek
philosophy. Aristotle (384--322 BC) is widely regarded as a foundational
thinker on this subject. In modern political economy {\em common good} is an
outcome of a collective action of a group of individuals that brings utility that each 
member of the group may enjoy. The utility is not assumed to be positive so the common good may turn out to be a {\em common bad}.
Common good is different from {\em public good} in which case all members of the society gain (or lose) utility. The common good can be {\em rivalrous} or not and {\em excludable} or not. This classification first appeared in the classical papers of Samuelson \cite{Samuelson1,Samuelson2}. 

In secret sharing the reconstruction of the secret can be done only collectively (by an authorised coalition) and knowledge of the secret brings utility for all or some of the group members so the secret qualifies to be a common good and Samuelson's classification is applicable. 

\begin{itemize}
\item The fact that the secret is {\em rivalrous} means that the less people
  know the secret the higher is the individual utility of those who know it.
\item The secret is {\em non-excludable} if all members of the group can (potentially) benefit from it but some, nevertheless, may miss out (say, if they come last). 
\end{itemize}

We give examples of common goods of various kinds:
\begin{itemize}
\item Rivalrous and excludable: buried treasure;
\item Rivalrous and non-excludable: wild fish, public transport;
\item Non-rivalrous and excludable: satellite television, wi-fi password;
\item Non-rivalrous and non-excludable: free-to-air television.
\end{itemize}

We note that, if the reference group coincides with the whole society, then common good becomes a {\em public good}.

\section{Halpern and Teague's Model}

Halpern and Teague and followers \cite{Halpern-Teague:2004,gordon2006rational,garay2013rational,kawachi2016general,kol2008cryptography} explored the rivalrous and excludable common good framework. 
 Given a run $r$ in the game tree, they introduce vector $\text{info}(r)$ to be the $n$-tuple $(\row tn)$, where $t_i=1$ if participant $i$ learns the secret and $t_i=0$, otherwise.  
They assume that utilities satisfy the following conditions:
\begin{itemize}
\item[U1.] If $\text{info}(r)=\text{info}(r')$, then $u_i(r)=u_i(r')$ for all $i\in [n]$.
\item[U2.] If $\text{info}_i(r)=1$ and $\text{info}_i(r')=0$, then $u_i(r)>u_i(r')$.
\item[U3.] If $\text{info}_i(r)=\text{info}_i(r')$, $\text{info}_j(r)\le \text{info}_j(r')$ for all $j\ne i$, and $\text{info}_{j_0}(r)< \text{info}_{j_0}(r')$ for some $j_0\in [n]$,  then $u_i(r)>u_i(r')$.
\end{itemize}
Condition U1 means that participants' utilities for all outcomes depend only on the information structure at the end of the game. U2 means that under any circumstances a participant prefers to know the secret to not knowing it. And U3 means that, if a participant learns the secret, she prefers fewer other people know it. 
It also convenient to normalise utilities in such a way that, if $\text{info}_i(r)=0$ for all $i\in [n]$, then $u_i(r)=0$ for all $i\in [n]$.  U3 means that if nobody learns the secret everybody's utility is zero but, if a participant does not learn the secret while another participant learns it, the utility of the former participant will be negative. 

 Shoham and  Tennenholtz~\cite{shoham2005non} in relation to non-cooperative computation  call the condition U2 {\em correctness}, which is the wish to compute the secret correctly. They also called U3 {\em exclusivity}, which is the wish that other agents do not compute the secret correctly. They assume a lexicographic ordering between these two, with correctness preceding exclusivity.\par
 
 Thus, conditions U1-U3 represent typical greedy utilities. We note that it is exactly condition U3 that makes the secret rivalrous and excludable. 


Halpern and Teague's model deals only with the classical Shamir's
$k$-out-of-$n$ access structures in which every coalition of $k$ participants
is authorised to know the secret while any smaller coalition is not authorised.
One of the main assumptions of the model is that at any node of the game tree all agents move simultaneously (this is called the synchronous framework). Thus, every node corresponds to what Gordon and Katz~\cite{gordon2006rational} call a communication round.   A move consists of disclosing to a \emph{subset} of other agents some of the information that a participant has. 
It is extremely important that Halpern and Teague assume that at any node the agents may only \emph{reveal} some information and receive it only after they made their revelation or declared her refusal to do so. The authors do not specify how the game tree is constructed and when the game ends. 

Also, they seem to assume private communication channels between agents (how
otherwise one can reveal their share only to one other agent). However this
assumption is dangerous.  Indeed, if under $2$-out-of-$3$ scheme, two of the
participants exchange their shares in private, then they would learn the secret
while the remaining participant will not. This is a Nash equilibrium.

 The class of protocols considered by Halpern and Teague  requires all participants to simultaneously disclose their shares. A participant can, however, not disclose her share deviating from the protocol. 
Most of the time the action of a single participant will be inconsequential. If at
most $k-2$ participants broadcast their shares, then nobody will know the secret
regardless of the action of our participant. If at least $k$ other participants broadcast their shares, then everybody will know the secret regardless of her action as well. An exceptional situation occurs when exactly $k-1$ other participants disclosed their shares of the secret and  all the rest do not. Then a participant is better to be in the abstaining group as then she and all the other $n-k-1$ abstaining participants, will know the secret but not those $k-1$ participants who disclosed their shares. As a result abstaining is a weakly dominant strategy and the only Nash equilibrium is when everybody abstains. 
 
Thus, the rational secret sharing scenario introduced by Halpern and Teague
\cite{Halpern-Teague:2004} represents some form of prisoner's dilemma when for
each member of the group all deterministic strategies except the strategy `to
abstain' are eliminated as weakly dominated by the strategy of abstaining. In
this sense it is rational for her not to participate in the recovery of the
secret as she may be hoping for a free ride and learn the secret without revealing her share.  We illustrate Halpern and Teague assumptions with the following examples.

\begin{description}

\item[Burried treasure.]
Suppose that there are $n$ inheritors of the fortune of a bequeather who has hidden a pot of gold in a secret location and created a $k$-out-of-$n$ secret sharing scheme dividing the secret into shares so that any $k$ inheritors can calculate the secret. At a specified date the inheritors must simultaneously broadcast their shares making them public or keeping them for themselves. Those who have learned the secret will then gather at the location of the buried treasure and divide the fortune equally between themselves.

\item[Secure Multiparty Computation.]
%
Secure Multiparty Computation (MPC) is a technology that allows you to compute on encrypted values maintaining privacy of the inputs. Using MPC a number of servers can jointly compute any function without learning the inputs to the function. Secret sharing is one cryptographic tool which allows you to do MPC by splitting inputs into shares and allow all computations be performed on shares. However, at the end,  parties end up with shares of the output and, if the output contains information leading to a rivalrous excludable common good, Halpern-Teague impossibility can play a role in impossibility to perform the final step of MPC. 

\item[Threshold cryptography: decrypting a message.]
Often the sender/receiver is not an individual but an organisation.  Shamir \cite{Shamir79} was the first to emphasise that a company's secret key for the purpose of digitally signing documents should not be given to a single individual. In such a case the secret decryption key of the organisation is split by a secret sharing scheme into shares and individuals of the organisation apply their shares on the incoming message. After which they send the results to a trusted combiner who reconstruct the message from partial results calculated by individual members \cite{Desmedt94ETT,Desmedt97JAIST}. It is clear that we cannot get rid of the trusted combiner since we will immediately find ourselves in the Halpern-Teague framework. 
\end{description}

What Halpern and Teague have shown is that in the case of $k$-out-of-$n$ access structure and deterministic protocol with simultaneous moves  the problem of free rider is unavoidable.

\begin{theorem}[Halpern and Teague, 2003]
For the $k$-out-of $n$ secret sharing scheme, if utilities of participants satisfy U1-U3, there is no deterministic synchronous protocol that will be a Nash equilibrium of the game surviving deletion of weakly dominated strategies such that, if this protocol is followed, at the end of the game at least one participant learns the secret.
\end{theorem}


Halpern and Teague strongly emphasised that they are working in synchronous
systems and they conjectured that in asynchronous systems there are no
practical mechanisms for secret sharing or multiparty computation satisfying
U1 - U3.  Below is a confirmation of their hypothesis. Of course, in such a
case the type of Nash equilibrium must be the subgame perfect Nash
equilibrium.  In this subsection we are still assuming U1-U3. 
Suppose now that:

\begin{itemize}
\item There is a prescribed order in which participants are supposed to be making moves;
\item Each move consists in the following: at the node where a participant has to move she is disclosing a subset of shares she knows to a subset (maybe empty) of other participants (and it is important that she herself does not get any new information); 
\item The game stops as soon as one of the participants learns the secret.
\end{itemize}

\begin{theorem}
Suppose that participants' utilities satisfy conditions U1 - U3. Then for any secret sharing scheme there is no asynchronous protocol that is a subgame perfect Nash equilibrium of the secret reconstruction and satisfies the following conditions:  
\begin{itemize} 
\item  if this protocol is followed, the game ends in finite time, and 
\item  at the end of the game at least one participant learns the secret.
\end{itemize}

\begin{proof}
 Finite games are usually solved by backward induction. Our game is not finite but still we can use the idea. Suppose that there is a protocol such that  if all participants follow it, the game ends with someone knowing the secret in finite time; and that it is a subgame perfect Nash equilibrium. Suppose that $r$ is a run in this game ending at a terminal node $e$. Due to the choice of utilities, some participants have positive payoffs (those who learned the secret) and some negative (those who did not learn it). The crux of the matter is that the participant who made the last move, let it be participant~$i$ at node $d$, cannot learn the secret since she only discloses information and cannot acquire it. As a result she gets a negative payoff. At the same time it was exactly the  information of participant $i$ that was sufficient for some participant to figure out the secret. So if participant $i$ at node $d$ discloses nothing, the game gets into a non-terminal node. Hence not a single participant has incentive to end the game. 
\end{proof}
\end{theorem}

\section{A New Model of Rational Secret Sharing}

Let us now deal with the non-rivalrous and non-excludable common good scenario
and support it by examples of situations which reflect this scenario.  In this
section we will assume that the secret sharing scheme has an arbitrary access structure $\Gamma$ with the set of minimal authorised coalitions $\Gamma_\text{min}$.

Firstly, in this scenario it is important whether or not the secret is recovered (or used) at the end of the game. If it was recovered, then the common good becomes available. Secondly, participation in the reconstruction may incur costs that parties have to pay in some form (be it the time spent or a burden of responsibility taken). Given a run $r$ in the game tree, we associate with it the following parameters
\[
\text{info}(r)=(t,{\bf s}),
\]
where $t=\text{info}_{1}(r) = 1$, if the secret was recovered, and $t=0$, if it was not; and ${\bf s}=(\row sn)$, where the value $s_i=\text{info}_{2,i}(r)=1$ means that agent $i$ participated in the recovery and $s_i=0$ means that she did not. We assume that utilities $\row un$ of parties satisfy the following conditions:
\begin{itemize}
\item[V1.] If $\text{info}_{1}(r)=1$, then $u_i(r)=N_i-c\cdot \text{info}_{2,i}(r)$ for all $i\in [n]$;
\item[V2.]  If $\text{info}_{1}(r)=0$, then  $u_i(r)=-c\cdot \text{info}_{2,i}(r)$.
\end{itemize}
where $N_i>0$ is the  value of the common good for party $i$ and $c>0$ is the cost of participation in the secret recovery.

Here are some examples. 

\begin{description}

\item[Authorising the project.] 
Imagine that a sufficiently authoritative group in a society, say a city
council, can launch (authorise) certain activity that will result in a public
good  which will be consumed by all members of the society bringing a
utility $N_i$ to member $i$ of the council (due to possible corruption the utility
$N_i$ of member $i$ could be much larger than utilities of other members).  To authorise the project one needs to
gather a coalition that would be authorised to learn (or use) the secret (say,
unlocking funds for the activity). The structure of authorised coalitions is given by an access structure $\Gamma$ of a secret sharing scheme, e.g., a 2/3 majority may be needed. The participation in the coalition, however, comes at a certain cost, say $c>0$, which may come in the form of time spent on negotiations or the responsibility for supporting the project. \par\smallskip

\item[Launching a nuclear missile.] In the former USSR any two of the three top state officials needed to activate their nuclear suitcases to launch a missile.  The recovery of the secret could open a possibility of creation of a public good (defeating an enemy) that all the society will consume collectively. It also gives us a sense of the possible cost that participants can pay for their participation in the recovery of the secret. Inevitably, many people will die as a result and resolving this moral dilemma may be daunting. This is, by the way, why maintaining the privacy of participants (i.e., who actually participated in the recovery) becomes paramount.\par\smallskip

\item[Threshold cryptography: signing a message.] If a message has to be signed on behalf of the organisation, say Microsoft, the signing key is split into shares between several authorised executives who sign the message with their shares and the combiner then transforms their results into the message signed by the organisation \cite{Desmedt94ETT,Desmedt97JAIST}. 
\end{description}

At this point we have to make several important observations.
\begin{enumerate}
\item We note that all three examples have a power sharing flavour. In all three, participants perform as experts approving or not approving the project in the first instance or a launch of a missile  in the second and a message in the third. Unlike Halpern-Teague scenario, in such a case non-reconstruction may be beneficial: parties do not reveal/use their shares if they suspect that their utilities may be negative as a result, i.e., the common good may be not good at all. As a side note, the common bad for the organisation may be actually a public good for the society (but this is not captured by our model).

\item In all three cases the secret itself is a meaningless combination of zeros and ones and knowledge of it has no value to participants (unless they want to engage in an illegal activity). An authorised coalition is not aiming to recover the secret but to {\em use} it to launch certain activity. In fact, the secret should never be recovered. 

\item It is amazing how different the two main applications of threshold cryptography are. The signing of a message on behalf of an organisation is totally different from decrypting an incoming message. Indeed, all incoming messages should be read, hence decrypted, while for signing a harmful message a person may be fired. Also there is no incentive to be the last signer. 
\end{enumerate}

We will show that the game with utilities V1,V2 has a natural set of Nash equilibria provided the cost of participation $c$ is not too high. Firstly, we show that the equilibria of such game are all in pure strategies.  A mixed strategy for participant $i$ is characterised by a single non-negative number $0\le \alpha_i\le 1$ which is a probability of participant $i$ disclosing her share in the recovery stage. 

Given a Nash equilibrium, we say that a player is {\em inessential} in this Nash equilibrium, if her utility does not depend on her strategy.

\begin{lemma}
In every Nash equilibrium for the game with utilities V1,V2 a player plays a pure strategy or else this player is inessential. 

\begin{proof}
The strategy of the $i$th member of the society is the probability $\alpha_i$ of participating in the recovery of the secret. Suppose that a vector of probabilities $\alpha^*=(\row {\alpha}n)$ is a Nash equilibrium. Suppose for some $i\in [n]$ we have $0<\alpha_i<1$. For convenience and without loss of generality we may assume that $i=n$. We look for the best responce of participant $n$ given the tuple of strategies $(\row {\alpha}{n-1})$, which we can therefore view as fixed. Then the expected utility of participant $n$, when she uses strategy $x$, would be 
\[
E_n(x)= x\cdot (-c)+ \left[x\cdot f_{\Gamma}(\row {\alpha}{n-1}) + g_{\Gamma}(\row {\alpha}{n-1})\right] N_n,
\]
where $f_{\Gamma}$ and $g_{\Gamma}$ be two functions that depend on the access structure $\Gamma$. Namely, the value $f_{\Gamma}(\row {\alpha}{n-1})$ is the probability of a non-authorised coalition which is a subset of $\{1,\ldots, n-1\}$ that together with $n$ becomes authorised; and $g_{\Gamma}(\row {\alpha}{n-1})$ is a probability of an authorised coalition in $\{1,\ldots, n-1\}$ without participation of $n$. 

As we see $E_n(x)$ is a linear function in $x$ and takes its extremal values either at 0 or at~1. This means pure strategies unless $x$ cancels out. In such a case player $n$ is inessential. This happens if $f_{\Gamma}(\row {\alpha}{n-1})N_n=c$.
\end{proof}
\end{lemma}

To illustrate the proof, here is the calculation of the expected utility of participant~3  in case $n=3$ and 2-out-of-3 scheme: 
\begin{align*}
E_3(x)&=x\cdot (-c) + \left[x( \alpha_1(1-\alpha_2)  + \alpha_2(1-\alpha_1))+ \alpha_1\alpha_2\right] N_3. 
\end{align*}
When $ [\alpha_1(1-\alpha_2)  + \alpha_2(1-\alpha_1)]N_3=c$ player 3 becomes inessential. This may be true for all three players. This, for example,  happens when $ 2\alpha(1-\alpha)N  =c$ and $N=N_1=N_2=N_3$, with all three players choose strategy $\alpha$.\par\medskip

Let $X\subseteq [n]$, then the characteristic vector $v_X$ of the subset $X$ is the vector for which $(v_X)_i=1$ in case $i\in X$ and  $(v_X)_i=0$, otherwise.

\begin{theorem}
Suppose we have a Nash equilibrium of the game with utilities satisfying V1 and V2 such that every player is essential.  Then this Nash equilibrium is one of the vectors $v_X$, where $X$ is the minimal authorised coalition of $\Gamma$ such that for every $i\in X$ we have $N_i>c$. If $\Gamma $ does not have a self-sufficient participant $i$ with $N_i>c$, then the zero vector is also a Nash equilibrium. All Nash equilibria survive deletion of weakly dominated strategies.
\begin{proof}
It is easy to see that for a partiy $i$ with $N_i<c$ the dominant strategy is always to abstain. By Lemma 1 we may assume that all values of a vector of Nash equilibrium are either 0 or
1. It is easy to see that the zero vector (in the absence of self-sufficient
participants) and  vectors $v_X$ for $X\in \Gamma_\text{min}$ with $N_i>c$ for every
$i\in X$ are Nash equilibria. Also vectors $v_Y$ for $Y\supset X$, where $X\in
\Gamma_\text{min}$ are not Nash equilibria since some participants may change their probabilities
to zero and save the amount $c>0$. If $X$ is not in authorised and somebody is playing a non-zero strategy, then
this participant can change their probabilities to zero and be better off. This proves the theorem.
\end{proof}
\end{theorem}

\section{Conclusion}
We have argued in this paper that secret sharing is used in different
circumstances and there are many realistic situations 
where the assumptions and conclusions of Halpern and Teague do not apply. 
It all depends on the nature of the secret, and, if it opens the way for consumption of a common good, 
another set of assumptions are needed leading to a range of non-trivial Nash equilibria. 

In particular, we show that there is a large set of scenarios, 
like the threshold cryptography and MPC, for which 
secrets should {\it not\/} be recovered but used. Indeed, in threshold signatures,
the signing key of the organisation should {\it not \/} be recovered.
If parties want to co-sign a document, they should {\it use\/} their
shares to do so. Similarly in MPC, the parties should {\it not\/} use their shares of
the inputs, but only the shares of the output.
Unlike Halpern and Teague scenario, non-participation may be beneficial and incentives for non-recovery in
such circumstances would make both threshold cryptography and MPC more attractive to potential users.

\newcommand{\speak}{\relax}\newcommand{\bibbf}{\relax}\newcommand{\bibpercent}[1]{}


\begin{thebibliography}{10}

\bibitem{Beerliova-TrubiniovaHirth08}
Z.~Beerliov{\'{a}}{-}Trub{\'{\i}}niov{\'{a}} and M.~Hirt.
\newblock Perfectly-secure {MPC} with linear communication complexity.
\newblock In R.~Canetti, editor, {\em Theory of Cryptography, Fifth Theory of
  Cryptography Conference, {TCC}}, volume 4948 of {\em Lecture Notes in
  Computer Science}, pp.~213--230. Springer, 2008.

\bibitem{BenOrGoldwasserWigderson88}
M.~Ben-Or, S.~Goldwasser, and A.~Wigderson.
\newblock Completeness theorems for non-cryptographic fault-tolerant
  distributed computation.
\newblock In {\em {Proceedings of the twentieth annual ACM Symposium Theory of
  Computing, STOC}}, pp.~1--10, {May 2--4,} 1988.

\bibitem{Blakley79}
G.~R. Blakley.
\newblock Safeguarding cryptographic keys.
\newblock In {\em Proceedings of the National Computer Conference},
  pp.~313--317, 1979.
\newblock vol.48.

\bibitem{Boyd86}
C.~Boyd.
\newblock Digital multisignatures.
\newblock In H.~Beker and F.~Piper, editors, {\em Cryptography and coding},
  pp.~241--246. Clarendon Press, 1989.
\newblock Royal Agricultural College, Cirencester, December 15--17, 1986.

\bibitem{CroftHarris86}
R.~A. Croft and S.~P. Harris.
\newblock Public-key cryptography and re-usable shared secrets.
\newblock In H.~Beker and F.~Piper, editors, {\em Cryptography and coding},
  pp.~189--201. Clarendon Press, 1989.
\newblock Royal Agricultural College, Cirencester, December 15--17, 1986.

\bibitem{Desmedt06BT}
Y.~Desmedt.
\newblock A high availability internetwork capable of accommodating compromised
  routers.
\newblock {\em BT Technology Journal}, 24(3), pp.~77--83, 2006.

\bibitem{DesmedtFrankel89Crypto}
Y.~Desmedt and Y.~Frankel\speak{}.
\newblock Threshold cryptosystems.
\newblock In G.~Brassard, editor, {\em Advances in Cryptology --- Crypto~'89,
  Proceedings (Lecture Notes in Computer Science 435)}, pp.~307--315.
  Springer-Verlag, 1990.
\newblock Santa Barbara, California, U.S.A., August 20--24\bibpercent{(48\%
  acceptance rate)}.

\bibitem{Desmedt94ETT}
Y.~G. Desmedt.
\newblock Threshold cryptography.
\newblock {\em European Transactions on Telecommunications}, 5(4),
  pp.~449--457, July-August 1994.
\newblock ({\bibbf Invited paper}).

\bibitem{Desmedt87}
Y.~Desmedt\speak{}.
\newblock Society and group oriented cryptography: a new concept.
\newblock In C.~Pomerance, editor, {\em Advances in Cryptology, Proceedings of
  Crypto~'87 (Lecture Notes in Computer Science 293)}, pp.~120--127.
  Springer-Verlag, 1988.
\newblock Santa Barbara, California, U.S.A., August 16--20.

\bibitem{Desmedt97JAIST}
Y.~Desmedt\speak{}.
\newblock Some recent research aspects of threshold cryptography.
\newblock In E.~Okamoto, G.~Davida, and M.~Mambo, editors, {\em Information
  Security, Proceedings (Lecture Notes in Computer Science 1396)},
  pp.~158--173. Springer-Verlag, 1997.
\newblock Invited lecture, September 17-19, 1997, Tatsunokuchi, Ishikawa,
  Japan, Springer-Verlag.

\bibitem{Desmedt05ITW}
Y.~Desmedt\speak{}.
\newblock Unconditionally private and reliable communication in an untrusted
  network.
\newblock In {\em IEEE Information Theory Workshop on Theory and Practice in
  Information-Theoretic Security, Proceedings}, pp.~38--41, October 16--19,
  2005.
\newblock Awaji Island, Japan.

\bibitem{DolevDworkWaartsYung93}
D.~Dolev, C.~Dwork, O.~Waarts, and M.~Yung.
\newblock Perfectly secure message transmission.
\newblock {\em Journal of the ACM}, 40(1), pp.~17--47, January 1993.

\bibitem{garay2013rational}
Juan Garay, Jonathan Katz, Ueli Maurer, Bjorn Tackmann, and Vassilis Zikas.
\newblock Rational protocol design: Cryptography against incentive-driven
  adversaries.
\newblock In {\em Foundations of Computer Science (FOCS), 2013 IEEE 54th Annual
  Symposium on}, pp.~648--657. IEEE, 2013.

\bibitem{GennaroRabinRabin98}
R.~Gennaro, M.~O. Rabin, and T.~Rabin.
\newblock Simplified {VSS} and fact-track multiparty computations with
  applications to threshold cryptography.
\newblock In {\em Proceedings of the Annual {ACM} Symposium on Principles of
  Distributed Computing ({PODC})}, pp.~101--111, 1998.

\bibitem{gordon2006rational}
S~Dov Gordon and Jonathan Katz.
\newblock Rational secret sharing, revisited.
\newblock In {\em SCN}, volume 4116, pp.~229--241. Springer, 2006.

\bibitem{Halpern-Teague:2004}
Joseph Halpern and Vanessa Teague.
\newblock Rational secret sharing and multiparty computation: Extended
  abstract.
\newblock In {\em Proceedings of the Thirty-sixth Annual ACM Symposium on
  Theory of Computing}, STOC '04, pp.~623--632, New York, NY, USA, 2004. ACM.

\bibitem{kawachi2016general}
Akinori Kawachi, Yoshio Okamoto, Keisuke Tanaka, and Kenji Yasunaga.
\newblock General constructions of rational secret sharing with expected
  constant-round reconstruction.
\newblock {\em The Computer Journal}, 60(5), pp.~711--728, 2016.

\bibitem{kol2008cryptography}
Gillat Kol and Moni Naor.
\newblock Cryptography and game theory: Designing protocols for exchanging
  information.
\newblock {\em Theory of Cryptography}, pp.~320--339, 2008.

\bibitem{KurosawaSuzuki09}
K.~Kurosawa and K.~Suzuki.
\newblock Truly efficient 2-round perfectly secure message transmission scheme.
\newblock {\em IEEE Transactions on Information Theory}, 55(11),
  pp.~5223--5232, 2009.

\bibitem{Liu68}
C.~L. Liu.
\newblock {\em Introduction to Combinatorial Mathematics}.
\newblock McGraw-Hill, New York, 1968.

\bibitem{McElieceSarwate81}
R.~J. McEliece and D.~V. Sarwate.
\newblock On sharing secrets and {Reed-Solomon} codes.
\newblock {\em Communications of the ACM}, 24(9), pp.~583--584, September 1981.

\bibitem{mishra2013cloud}
Ankur Mishra, Ruchita Mathur, Shishir Jain, and Jitendra~Singh Rathore.
\newblock Cloud computing security.
\newblock {\em International Journal on Recent and Innovation Trends in
  Computing and Communication}, 1(1), pp.~36--39, 2013.

\bibitem{socio-rational}
Mehrdad Nojoumian and Douglas~R. Stinson.
\newblock Socio-rational secret sharing as a new direction in rational
  cryptography.
\newblock In Jens Grossklags and Jean Walrand, editors, {\em Decision and Game
  Theory for Security}, pp.~18--37, Berlin, Heidelberg, 2012. Springer Berlin
  Heidelberg.

\bibitem{Samuelson1}
Paul~A. Samuelson.
\newblock The pure theory of public expenditure.
\newblock {\em The Review of Economics and Statistics}, 36(4), pp.~387--389,
  1954.

\bibitem{Samuelson2}
Paul~A. Samuelson.
\newblock Diagrammatic exposition of a theory of public expenditure.
\newblock {\em The Review of Economics and Statistics}, 37(4), pp.~350--356,
  1955.

\bibitem{Shamir79}
A.~Shamir.
\newblock How to share a secret.
\newblock {\em Communications of the ACM}, 22, pp.~612--613, November 1979.

\bibitem{shoham2005non}
Yoav Shoham and Moshe Tennenholtz.
\newblock Non-cooperative computation: Boolean functions with correctness and
  exclusivity.
\newblock {\em Theoretical Computer Science}, 343(1), pp.~97--113, 2005.

\bibitem{Yao82a}
A.~C. Yao.
\newblock Protocols for secure computations.
\newblock In {\em {23rd Annual Symposium on Foundations of Computer Science
  (FOCS)}}, pp.~160--164. IEEE Computer Society Press, 1982.

\bibitem{Yao86}
A.~C. Yao.
\newblock How to generate and exchange secrets.
\newblock In {\em {27th Annual Symposium on Foundations of Computer Science
  (FOCS)}}, pp.~162--167. IEEE Computer Society Press, 1986.
\newblock Toronto, Ontario, Canada, October 27--29, 1986.

\end{thebibliography}
\end{document}